\documentclass[11pt,english]{article}
\usepackage[T1]{fontenc}
\usepackage[latin9]{inputenc}
\usepackage{geometry}
\usepackage{float}
\usepackage{graphicx}
\usepackage{amssymb}

\makeatletter

\providecommand{\tabularnewline}{\\}

\newcommand{\lyxaddress}[1]{
\par {\raggedright #1
\vspace{1.4em}
\noindent\par}
}

\usepackage{a4wide}
\usepackage{a4wide}

\usepackage{babel}
\makeatother

\begin{document}

\title{Solving topological defects via fusion }

\author{Z. Bajnok and Zs. Simon}

\maketitle

\lyxaddress{\begin{center}
\emph{Theoretical Physics Research Group of the Hungarian Academy
of Sciences,}\\
\emph{ H-1117 P\'azm\'any s. 1/A, Budapest, Hungary}
\par\end{center}}

\begin{abstract}
Integrable defects in two-dimensional integrable models are purely
transmitting thus topological. By fusing them to integrable boundaries
new integrable boundary conditions can be generated, and, from the
comparison of the two solved boundary theories, explicit solutions
of defect models can be extracted. This idea is used to determine
the transmission factors and defect energies of topological defects
in sinh-Gordon and Lee-Yang models. The transmission factors are checked
in Lagrangian perturbation theory in the sinh-Gordon case, while the
defect energies are checked against defect thermodynamic Bethe ansatz
equations derived to describe the ground-state energy of diagonal
defect systems on a cylinder. Defect bootstrap equations are also
analyzed and are closed by determining the spectrum of defect bound-states
in the Lee-Yang model. 
\end{abstract}

\section{Introduction}

Recently, there has been an increasing interest in integrable quantum
field theories including defects or impurities. This is motivated
both by the realistic physical applications in statistical and solid
state physics and also by the need of theoretical understanding of
this so-far unexplored field. 

The community of integrable systems have not payed much attention
to defect theories at the beginning due to the no-go theorem formulated
by Delfino, Mussardo and Simonetti in \cite{DMS1,DMS2}. The theorem,
formulated originally for diagonal theories and extended later for
a large class of non-diagonal ones in \cite{Defboot}, states that
a relativistically invariant theory with a non-free integrable interaction
in the bulk can allow only two types of integrable defects: the purely
reflecting and the purely transmitting ones. (Although some effort
has been made to overcome this obstacle by giving up Lorentz invariance,
see for instance \cite{CMRS} and references therein, in the present
paper we restrict ourselves to the relativistically invariant case.) 

The analysis of boundary integrable theories was initiated in \cite{GZ}
by formulating, in an axiomatic way, the properties of the reflection
matrix: unitarity, boundary crossing unitarity and boundary bootstrap
equation. The boundary bootstrap framework was completed by introducing
boundary Coleman-Thun mechanism \cite{BCT} and the bulk bootstrap
equations \cite{FK}. Later this framework got a sound basis by developing
boundary quantum field theories from first principles in \cite{BRF,BBT}.
The success of the boundary bootstrap approach resulted in a large
class of closed bootstrap theories in which the boundary reflection
factors together with the spectrum of boundary excited states were
determined \cite{GZ,BCT,Valentina,CT,MD,BPTT,TGZS}. The solutions,
obtained by the bootstrap method, are not connected, however, to other
formulations of the model such as to the classical field theory or
to the perturbed conformal field theory (both defined by a Lagrangian).
To connect the different descriptions one either has to check the
reflection factors perturbatively, like in \cite{Edpert}, or solve
the theories in finite volume. The boundary thermodynamic Bethe ansatz
(BTBA), developed in \cite{LMSS}, systematically sums up the finite
size corrections by taking into account the scatterings and reflections.
By analyzing its small volume limit the needed link between the bootstrap
and perturbed conformal field theoretical descriptions can be established. 

As the no-go theorem showed non-free integrable defect theories are
purely transmitting. This fact kept back the researchers for some
time to analyze these models until new life was put into the subject
due to their explicit Lagrangian realizations \cite{ClDef}. Following
the original idea many integrable defect theories were constructed
at the classical level \cite{DAT,GYZ1,Vincent}. The basis for the
quantum formulation of defect theories is provided by the folding
trick \cite{DefBound} by which one can map any defect theory into
a boundary one. As a consequence defect unitarity, defect crossing
symmetry and defect bootstrap equations together with defect Coleman-Thun
mechanism are derived. Despite of these results the explicitly solved
relativistically invariant defect quantum field theories are quite
rare, containing basically the sine-Gordon and affine Toda field theories
\cite{KL,sgdef,QAT} and even in these cases the explicit relation
to their Lagrangian have not been worked out yet. 

One may think that purely transmitting theories are too simple and
there is no point to analyze them, but we would like to argue that
they carry very important information about an integrable quantum
field theory, without which our knowledge cannot be complete. Purely
transitivity implies the conservation of momentum from which the topological
nature of the defect follows. Thus, such defects can freely be transported
in space without affecting the physics of the theory. We can either
move them close to each other or move them to integrable boundary
conditions and, as a result, new integrable boundary conditions can
be generated. These ideas were successfully applied in conformal field
theories \cite{PZ,GW,Ingo}, in integrable lattice models \cite{BP,CMOP}
and the aim of the present paper is to exploit it in solving integrable
defects in the sinh-Gordon and Lee-Yang theories. 

The paper is organized as follows: In section 2, on the example of
the sinh-Gordon theory, we show how new boundary conditions can be
obtained by fusing integrable defects to boundaries at the classical
level. We focus on two cases in detail: fusing the integrable defect
to Dirichlet boundary condition (DBC) the perturbed Neumann boundary
condition (PNBC) can be obtained, while fusing it to the Neumann BC
a new class of time-dependent integrable BCs can be generated. Section
3 recalls the quantum version of the fusion method together with the
properties of transmission factors. By comparing the already known
reflection factors and boundary energies of the DBC to those of the
PNBC solutions for the defect transmission factor and defect energy
can be extracted. The same method is then used to determine defect
energies and transmission factors of the scaling Lee-Yang model. This
method not only provides the explicit solutions of the sinh-Gordon
defect theory but also relates its parameter to that of the Lagrangian.
Since the relation obtained here is different from the suggestion
of \cite{sgdef} we perform a perturbative analysis at one-loop level
in section 4. (In the subsequent paper \cite{QAT} the same authors
raised the possibility of the quantum renormalization of the transmission
parameter, which is confirmed at one-loop level here). In section
5 the calculated defect energies are subject to another consistency
check. For this we derive a TBA equation to describe the groundstate
energy of a diagonal defect system on a cylinder. From a careful ultra-violet
(UV) analysis defect energies are extracted and the previous results
are verified. By completing the defect bootstrap program we analyze
the singularity structure of the transmission factors in section 6.
Since the sinh-Gordon transmission factor does not contain any singularity
in the physical strip we analyze the Lee-Yang model only. For each
pole of the transmission factor in the physical strip we associate
either a defect boundstate or a defect Coleman-Thun diagram and calculate
the excited transmission factors in the former case from the defect
bootstrap equation. Finally, we conclude in section 7 and give directions
for further research.

\section{Fusion method at the Lagrangian level\label{sec:Lagrangianshg}}

In this section we demonstrate, on the example of the sinh-Gordon
(ShG) theory, how new integrable boundary conditions can be obtained
by the fusion method at the Lagrangian or classical level \cite{Peter}. 

The ShG theory is defined on the whole line by the following Lagrangian\begin{equation}
\mathcal{L}_{\mathrm{ShG}}(\Phi)=\frac{1}{2}(\partial_{t}\Phi)^{2}-\frac{1}{2}(\partial_{x}\Phi)^{2}-\frac{m_{\mathrm{cl}}^{2}}{b^{2}}\cosh b\Phi\,\,\quad,\label{ShGbulkL}\end{equation}
It describes an integrable field theory and by restricting the theory
to the half line this property can be only maintained, whenever the
following boundary potential is introduced \cite{GZ}\begin{equation}
\mathcal{L}_{\mathrm{BShG}}=\Theta(-x)\mathcal{L}_{\mathrm{ShG}}(\Phi)-\delta(x)B(\Phi)\quad;\qquad B(\Phi)=M_{0}\cosh\frac{b}{2}(\Phi-\varphi_{0})\label{eq:shGBC}\end{equation}
By varying $M_{0}$ from $0$ to $\infty$ the arising boundary condition
interpolates between the Neumann $\partial_{x}\Phi\vert_{x=0}=0$
and the Dirichlet $\Phi(0,t)=\varphi_{0}$ ones. 

The most general integrable defect condition can be obtained by the
analytical continuation of the sine-Gordon result \cite{sgdef}. The
Lagrangian in the ShG case reads as \begin{equation}
\mathcal{L}_{\mathrm{DShG}}=\Theta(-x)\mathcal{L}_{\mathrm{ShG}}(\Phi_{-})-\delta(x)D(\Phi_{-},\Phi_{+})+\Theta(x)\mathcal{L}_{\mathrm{ShG}}(\Phi_{+})\label{eq:ShGdefL}\end{equation}
where the defect potential contains just one single parameter: \begin{eqnarray*}
2D(\Phi_{-},\Phi_{+}) & = & \Phi_{+}\dot{\Phi}_{-}-\Phi_{-}\dot{\Phi}_{+}+M_{\mathrm{cl}}e^{\mu}\cosh\frac{b}{2}(\Phi_{+}+\Phi_{-})+M_{\mathrm{cl}}e^{-\mu}\cosh\frac{b}{2}(\Phi_{+}-\Phi_{-})\end{eqnarray*}
Here $\Phi_{\mp}$ are the fields living on the left/right half-line,
respectively and $M_{\mathrm{cl}}=\frac{4m_{\mathrm{cl}}}{b^{2}}$. 

The fusion idea is based on the integrability of the defect: Integrability
guaranties the existence of an infinite number of commuting conserved
charges which results in the possibility of shifting the trajectories
of particles, without changing the amplitude of any scattering process.
The shifting of all the trajectories can alternatively be described
by shifting the location of the defect, which then, does not alter
the physics.

At the level of the Lagrangian this observation can be formulated
in the following way: The spectrum of the system, which contains a
defect in the origin, $x=0$, in front of a boundary, located at $x=a$,
\[
\mathcal{L}_{\mathrm{DBShG}}=\Theta(-x)\mathcal{L}_{\mathrm{ShG}}(\Phi_{-})-\delta(x)D(\Phi_{-},\Phi_{+})+\Theta(x)\Theta(a-x)\mathcal{L}_{\mathrm{ShG}}(\Phi_{+})-\delta(x-a)B(\Phi_{+})\]
does not actually depend on $a$. Thus we can perform the $a\to0$
limit and represent the same system as a boundary one, but with a
different (dressed) boundary condition:\[
\mathcal{L}_{\mathrm{DBShG}}=\Theta(-x)\mathcal{L}_{\mathrm{ShG}}(\Phi_{-})-\delta(x)B^{'}(\Phi_{-},\Phi_{+})\quad;\qquad B^{'}(\Phi_{-},\Phi_{+})=D(\Phi_{-},\Phi_{+})+B(\Phi_{+})\]
The field $\Phi_{+}$ lives on the boundary only and can be thought
naively to be a boundary degree of freedom. It does not have, however,
any kinetic term so it merely implements a new time-dependent integrable
boundary condition. Let us specify these findings in two concrete
examples that will be used later on. 

If the original boundary condition is the Dirichlet one with $\Phi(a,t)=\phi_{0}$,
then the arising dressed boundary condition is\begin{equation}
B^{'}(\Phi_{-},\Phi_{+})=D(\Phi_{-},\varphi_{0})=\frac{M_{\mathrm{cl}}e^{\mu}}{2}\cosh\frac{b}{2}(\Phi_{-}+\phi_{0})+\frac{M_{\mathrm{cl}}e^{-\mu}}{2}\cosh\frac{b}{2}(\Phi_{-}-\phi_{0})\label{eq:DNfusion}\end{equation}
where we dropped the total time derivatives. This BC is exactly of
the form of (\ref{eq:shGBC}) with parameters \[
M_{\mathrm{cl}}\cosh(\mu\pm\frac{b}{2}\phi_{0})=M_{0}e^{\mp\frac{b}{2}\varphi_{0}}\]
Thus by fusing the integrable defect to the DBC we can reconstruct
the most general (two parameter family of) PNBCs. Interestingly $\phi_{0}$
and $\mu$ are the classical analogues of the parameters in which
the boundary reflection is factorized \cite{BRZ}, see also (\ref{eq:RpN},\ref{eq:UVIRpN})
in section 3.

By fusing the defect to the NBC we obtain the boundary potential\[
B^{'}(\Phi_{-},\Phi_{+})=D(\Phi_{-},\Phi_{+})\]
Variation of action provides BCs in the form:\begin{equation}
\partial_{t}\Phi_{-}\vert_{x=0}=-\frac{\partial B^{'}(\Phi_{+},\Phi_{-})}{\partial\Phi_{+}}\quad;\qquad(\partial_{t}\Phi_{+}-\partial_{x}\Phi_{-})\vert_{x=0}=\frac{\partial B^{'}(\Phi_{+},\Phi_{-})}{\partial\Phi_{-}}\label{eq:Nfusion}\end{equation}
By expressing $\Phi_{+}$ in terms of $\Phi_{-}$ and $\partial_{t}\Phi_{-}$
then plugging back to the second equation we obtain a highly non-trivial
boundary condition for $\Phi_{-}$ containing its second time derivative,
which is nevertheless integrable as it follows from the construction.
Obviously, this solution was not covered by the two parameter family
of (time-independent) integrable boundary conditions determined in
\cite{GZ}, thus by the fusion method we were able to construct a
new type of integrable BC. By fusing other integrable defects with
new free parameters to this dressed boundary we can generate integrable
BCs with as many parameters as we want. What is nice in the construction,
that the solution of the defects transmission factor will provide,
via the fusion method, solutions for these general integrable BCs,
too, as we will show in the next section.

Finally, we note that similar construction can be used in the case
of the Lee-Yang model, however, the explicit form of the integrable
defect perturbation has not been identified at the Lagrangian level
yet ( for details see the next section).

\section{Fusion method in the bootstrap}

For simplicity we present the fusion idea in the case of an integrable
diagonal scattering theory with one particle type of mass $m$. The
general discussion can be found in \cite{DefBound}. 

In integrable bulk theories multi-particle scattering processes factorize
into the product of two particle scatterings, $S(\theta_{12})$, where
$\theta_{12}=\theta_{1}-\theta_{2}$ is the rapidity difference of
the scattering particles whose momenta are parametrized as $p_{i}=m\sinh\theta_{i}$.
In relativistically invariant theories the two particle $S$-matrix
satisfies unitarity and crossing symmetry \[
S(-\theta)=S^{-1}(\theta)\quad;\qquad S(i\pi-\theta)=S(\theta)\]

Once boundaries are introduced the basic process is the multi-particle
reflection. Integrability ensures its factorization into pairwise
scatterings $S(\theta_{ij})$ and individual reflections $R(\theta_{i})$,
where $\theta_{i}$ is the rapidity of the reflected particle. The
reflection matrix satisfies unitarity and boundary crossing unitarity
\cite{GZ}\[
R(-\theta)=R^{-1}(\theta)\quad;\qquad R(\frac{i\pi}{2}-\theta)=S(2\theta)R(\frac{i\pi}{2}+\theta)\]

Integrable non-free defects are severely restricted: they are either
purely reflecting (thus boundaries, like above) or purely transmitting.
The latter case can be described by the two, left ($-$) and right
($+$), transmission matrices $T_{-}(\theta)$ and $T_{+}(-\theta)$.
We parametrize $T_{+}$ such a way that for its physical domain ($\theta<0$)
its argument is always positive. Transmission factors satisfy unitarity
and defect crossing symmetry \cite{DefBound}\begin{equation}
T_{+}(-\theta)=T_{-}^{-1}(\theta)\quad;\qquad T_{-}(\theta)=T_{+}(i\pi-\theta)\label{Defprop}\end{equation}

If we place a defect with transmission matrices $T_{\pm}(\theta)$
in front of a boundary with reflection matrix $R(\theta)$ then the
fused boundary will also be integrable and have reflection factor
$R^{'}(\theta)$:

\begin{equation}
R^{'}(\theta)=T_{+}(\theta)R(\theta)T_{-}(\theta)\label{eq:Qfusion}\end{equation}
The correspondence (\ref{eq:Qfusion}) between the original $R(\theta)$
and the fused $R^{'}(\theta)$ reflection factors can be used either
to generate new BCs or, if the two BCs are already known, to solve
defect transmission factors. This will be illustrated in the next
subsections for the sinh-Gordon and Lee-Yang models.

\subsection{Solution of defect sinh-Gordon theory }

The spectrum of ShG theory defined by (\ref{ShGbulkL}) consists of
one particle type with mass \cite{masscale} \begin{equation}
m=\frac{4\sqrt{\pi}}{\Gamma(\frac{1-B}{2})\Gamma(1+\frac{B}{2})}\left(\frac{-\pi m_{\mathrm{cl}}^{2}\Gamma(1+b^{2})}{b^{2}\Gamma(-b^{2})}\right)^{\frac{1}{2+2b^{2}}}\qquad;\qquad B=\frac{b^{2}}{8\pi+b^{2}}\label{eq:mandB}\end{equation}
 The two particle scattering matrix is given by \[
S=\frac{\sinh\theta-i\sin B\pi}{\sinh\theta+i\sin B\pi}=-(-B)(1+B)\quad,\qquad(x)=\frac{\sinh(\frac{\theta}{2}+\frac{i\pi x}{2})}{\sinh(\frac{\theta}{2}-\frac{i\pi x}{2})}\]
 It has no poles in the physical strip: $0\leq\theta<i\pi$ and is
invariant under the weak-strong duality, $\frac{b^{2}}{8\pi}\rightarrow\frac{8\pi}{b^{2}}$.
The bulk energy density turns out to be \cite{Ebulk} \begin{equation}
\epsilon_{\mathrm{bulk}}=\frac{m^{2}}{8\sinh\pi B}\label{eq:becshg}\end{equation}

Integrable boundary conditions can be either Dirichlet type with $\Phi(0,t)=\phi_{0}$
or PN type (\ref{eq:shGBC}). The corresponding reflection factors
can be obtained from the analytical continuation of the sine-Gordon's
first breather's one \cite{GZ,Ghoshal}. In the Dirichlet case it
reads as 

\begin{equation}
R_{\mathrm{Dir}}(\theta,\eta_{\mathrm{Dir}})=\frac{\left(\frac{1}{2}\right)\left(1-\frac{B}{2}\right)}{\left(\frac{3}{2}-\frac{B}{2}\right)}\frac{\left(\frac{iB\eta_{\mathrm{Dir}}}{\pi}-\frac{1}{2}\right)}{\left(\frac{iB\eta_{\mathrm{Dir}}}{\pi}+\frac{1}{2}\right)}\label{eq:RDir}\end{equation}
where the reflection parameter $\eta_{\mathrm{Dir}}$ is related to
$\phi_{0}$ as \begin{equation}
\eta_{\mathrm{Dir}}=\frac{4\pi}{b}\phi_{0}\label{eq:UVIRDir}\end{equation}
 The boundary energy has been also calculated \cite{LMSS}\begin{equation}
\epsilon_{\mathrm{bdry}}^{\mathrm{Dir}}(\eta_{\mathrm{Dir}})=\frac{m}{4\sin B\pi}\left(2\cosh B\eta_{\mathrm{Dir}}-\sin\frac{\pi B}{2}-\cos\frac{\pi B}{2}-1\right)\label{eq:EDir}\end{equation}
In the PN case the reflection factor turns out to be \begin{equation}
R_{\mathrm{PN}}(\theta,\eta,\vartheta)=\frac{\left(\frac{1}{2}\right)\left(1-\frac{B}{2}\right)}{\left(\frac{3}{2}-\frac{B}{2}\right)}\frac{\left(\frac{iB\eta}{\pi}-\frac{1}{2}\right)}{\left(\frac{iB\eta}{\pi}+\frac{1}{2}\right)}\frac{\left(\frac{iB\vartheta}{\pi}-\frac{1}{2}\right)}{\left(\frac{iB\vartheta}{\pi}+\frac{1}{2}\right)}\label{eq:RpN}\end{equation}
while the relation of $\eta,\vartheta$ to the parameters of the Lagrangian
(UV-IR relation) is \cite{BRZ}\begin{equation}
M\cosh\frac{b^{2}}{8\pi}(\eta\pm\vartheta)=M_{0}e^{\mp\frac{b\varphi_{0}}{2}}\quad;\qquad M=m_{\mathrm{cl}}\sqrt{\frac{2}{b^{2}\sin(b^{2}/8)}}\label{eq:UVIRpN}\end{equation}
The boundary energy has been also determined \cite{BUVIR,BRZ} as
\begin{equation}
\epsilon_{\mathrm{bdry}}^{\mathrm{PN}}(\eta,\vartheta)=\frac{m}{4\sin B\pi}\left(2\cosh B\eta+2\cosh B\vartheta-\sin\frac{\pi B}{2}-\cos\frac{\pi B}{2}-1\right)\label{eq:EpN}\end{equation}
We note that the results - both for the UV-IR relation and for the
boundary energy - were obtained in the framework of perturbed BCFT
in which the perturbing operator is normal-ordered to have a definite
scaling dimension. 

The integrable defect potential for the sinh-Gordon model can be written
as in (\ref{eq:ShGdefL}). Let us denote the transmission factors
by $T_{\pm}(\theta,\mu)$. Fusing classically this defect to a DBC
a PNBC can be obtained (\ref{eq:DNfusion}). The quantum analogue
of this statement in view of (\ref{eq:Qfusion}) is \[
R_{\mathrm{PN}}(\theta,\eta,\vartheta)=T_{+}(\theta,\mu)R_{\mathrm{Dir}}(\theta,\eta_{\mathrm{Dir}})T_{-}(\theta,\mu)\]
Comparing the reflection factor of the PNBC (\ref{eq:RpN}) to that
of the Dirichlet one (\ref{eq:RDir}) and taking into account defect
unitarity and defect crossing symmetry (\ref{Defprop}) we can extract
the transmission factors for the defect. The simplest possible solution
corresponds to $\eta=\eta_{\mathrm{Dir}}$ and

\begin{equation}
T_{-}(\theta)=-i\frac{\sinh\left(\frac{\theta}{2}-\frac{i\pi}{4}+\frac{B\vartheta}{2}\right)}{\sinh\left(\frac{\theta}{2}+\frac{i\pi}{4}+\frac{B\vartheta}{2}\right)}\quad;\qquad T_{+}(\theta)=i\frac{\sinh\left(\frac{\theta}{2}-\frac{i\pi}{4}-\frac{B\vartheta}{2}\right)}{\sinh\left(\frac{\theta}{2}+\frac{i\pi}{4}-\frac{B\vartheta}{2}\right)}\label{Defsol}\end{equation}
All other solutions contain additional CDD type factors satisfying
(\ref{Defprop}). Actually the solution (\ref{Defsol}) itself is
a CDD factor, therefore it is the simplest non-trivial solution of
(\ref{Defprop}).

To find the correspondence between the parameter of the quantum transmission
factor $\vartheta$ and the Lagrangian parameter $\mu$ we follow
the following strategy: Since the boundary results are derived in
the perturbed BCFT normalization we allow not only the parameter $\mu$
but also $M_{{\rm \mathrm{cl}}}$ to renormalize. Their renormalized
quantum values are determined from the requirement that when fusing
the defect to the DBC with (\ref{eq:UVIRDir}) we obtain the PNBC
with (\ref{eq:UVIRpN}). The unique solution turns out to be \begin{equation}
Me^{\pm\frac{b^{2}\vartheta}{8\pi}}=M_{\mathrm{cl}}e^{\pm\mu}\label{eq:UVIRdef}\end{equation}
The renormalization of $M_{\mathrm{cl}}$ may depend on the scheme
in which the quantum potential is defined. The $b\to0$ limit, in
which $M\to M_{\mathrm{cl}}$, shows that $\vartheta$ is the quantum
renormalized version of $\mu$. 

Also the defect energy can be extracted as the difference of the boundary
energies corresponding to the PN (\ref{eq:EpN}) and to the Dirichlet
(\ref{eq:EDir}) one: \begin{equation}
\epsilon_{\mathrm{Def}}(\vartheta)=\epsilon_{\mathrm{bdry}}^{\mathrm{PN}}(\eta,\vartheta)-\epsilon_{\mathrm{bdry}}^{\mathrm{Dir}}(\eta)=\frac{m\cosh B\vartheta}{2\sin B\pi}\label{eq:Edef}\end{equation}
Summarizing, by the fusion method we were able to solve the defect
theory defined by the Lagrangian (\ref{eq:ShGdefL}): The transmission
factors are (\ref{Defsol}), the defect energy is (\ref{eq:Edef}),
and the bootstrap parameter $\vartheta$ parametrizes the Lagrangian
as (\ref{eq:UVIRdef}). We spend the next section to provide consistency
checks of this solution.

Once the defect theory is solved we can use it to generate new integrable
BCs from known ones. In the example presented in section \ref{sec:Lagrangianshg}
the defect with parameter $\mu$ was fused to the NBC to generate
a more general integrable BC (\ref{eq:Nfusion}). The quantum version
of this fusion dresses up the Neumann reflection factor \begin{equation}
R_{\mathrm{N}}(\theta)=\frac{\left(\frac{1}{2}\right)\left(1-\frac{B}{2}\right)}{\left(\frac{3}{2}-\frac{B}{2}\right)}\frac{\left(\frac{1}{2}-\frac{B}{2}\right)}{\left(\frac{1}{2}+\frac{B}{2}\right)}\label{eq:RNeumann}\end{equation}
to the reflection factor\begin{equation}
R(\theta,\vartheta)=T_{+}(\theta,\mu(\vartheta))R_{\mathrm{N}}(\theta)T_{-}(\theta,\mu(\vartheta))=\frac{\left(\frac{1}{2}\right)\left(1-\frac{B}{2}\right)}{\left(\frac{3}{2}-\frac{B}{2}\right)}\frac{\left(\frac{1}{2}-\frac{B}{2}\right)}{\left(\frac{1}{2}+\frac{B}{2}\right)}\frac{\left(\frac{iB\vartheta}{\pi}-\frac{1}{2}\right)}{\left(\frac{iB\vartheta}{\pi}+\frac{1}{2}\right)}\label{eq:RNdressed}\end{equation}
Thus we solved the more general integrable BC defined by (\ref{eq:Nfusion})
without doing any serious calculation. It is important to note, that
the extra factor appearing in (\ref{eq:RNdressed}) compared to (\ref{eq:RNeumann})
is a CDD factor. Consequently, we have determined the physical meaning
of the CDD factors appearing in the reflection factors: they represent
integrable defects of the form of (\ref{eq:Nfusion}) standing in
front of integrable boundaries. In principle, we can fuse as many
integrable defects with various parameters as we want, the resulting
theory can be solved and its reflection factor contains the corresponding
boundary CDD factors. 

Finally, we note that placing two defects with parameters $\vartheta_{\pm}=\pm i(1-\frac{\pi}{2B})$
after each other both the scattering matrix and the energy of a standing
particle can be reproduced. Thus the defect with imaginary parameter
can be considered as a 'half' particle. Similar phenomena was observed
in \cite{sgdef} at the classical level.

\subsection{Solution of defect scaling Lee-Yang model}

The scaling Lee-Yang model can be defined as the perturbation of the
$\mathcal{M}_{(2,5)}$ conformal minimal model with central charge
$c=-\frac{22}{5}$. It contains two modules of the Virasoro algebra
corresponding to the $Id$ and the $\varphi(z,\bar{z})$ primary fields
with weight $(0,0)$ and $(-\frac{1}{5},-\frac{1}{5})$, respectively.
The only relevant perturbation by the field $\varphi$ results in
the simplest scattering theory with one neutral particle of mass $m$
and scattering matrix \cite{SYLbulk} \[
S(\theta)=\frac{\sinh\theta+i\sin\frac{\pi}{3}}{\sinh\theta-i\sin\frac{\pi}{3}}=-\left(\frac{1}{3}\right)\left(\frac{2}{3}\right)\]
The pole at $\theta=\frac{i\pi}{3}$ shows that the particle can form
a bound-state. The relation\[
S(\theta+i\frac{\pi}{3})S(\theta-i\frac{\pi}{3})=S(\theta)\]
however, implies that the bound-state is the original particle itself
and the bulk bootstrap is closed. The bulk energy constant is given
by $\epsilon_{\mathrm{bulk}}=-\frac{1}{4\sqrt{3}}m^{2}$.

We can impose two conformal invariant boundary conditions in the model
\cite{BYLTBA,BYL1pt}. They can be labeled by $\mathbb{I}$ and $\Phi$
and correspond to the highest weight representations of a single copy
of the Virasoro algebra with weight $0$ and $-\frac{1}{5}$, respectively.
Introducing the integrable bulk perturbations with the $\mathbb{I}$
conformal invariant boundary condition the integrability is maintained
and the reflection factor of the particle can be written as \[
R_{\mathbb{I}}(\theta)=\left(\frac{1}{2}\right)\left(\frac{1}{6}\right)\left(-\frac{2}{3}\right)\]
while the boundary energy is given by $\epsilon_{\mathrm{bdry}}^{\mathbb{I}}=\frac{m}{2}\left(\sqrt{3}-1\right)$.
If the conformal invariant boundary condition corresponds to $\Phi$
then additionally to the bulk perturbation we can introduce a one-parameter
family of integrable boundary perturbations and the corresponding
reflection factor turns out to be\[
R_{b}(\theta)=\left(\frac{1}{2}\right)\left(\frac{1}{6}\right)\left(-\frac{2}{3}\right)\left(\frac{b-1}{6}\right)\left(\frac{b+1}{6}\right)\left(\frac{5-b}{6}\right)\left(\frac{-5-b}{6}\right)\]
while the boundary energy is $\epsilon_{\mathrm{bdry}}=\frac{m}{2}\left(\sqrt{3}-1+2\sin\frac{b\pi}{6}\right)$.
The boundary bound-states were analyzed in \cite{BCT} where the boundary
bootstrap program was carried out. 

The Lee-Yang model has two types of conformal defects \cite{YLD},
but only one of them admits relevant chiral defect fields. They have
weights $(-\frac{1}{5},0)$, $(0,-\frac{1}{5})$. We conjecture that
perturbing in the bulk and with a certain combination of these defect
fields we can maintain integrability and arrive at a purely transmitting
theory. We plan to analyze this issue systematically in a forthcoming
publication. Let us denote the transmission factors of this integrable
defect by $T_{\pm}(\theta)$. Using that the fusion of the defect
to the $\mathbb{I}$ boundary results in the perturbed $\Phi$ boundary
we have\[
R_{b}(\theta)=T_{+}(\theta,b)R_{\mathbb{I}}(\theta)T_{-}(\theta,b)\]
This is supported by the fact that fusing the conformal defect to
the $\mathbb{I}$ boundary we obtain the $\Phi$ boundary. Since the
particle appears as a bound-state in the two particle scattering process
the transmission matrix satisfies the defect bootstrap equation \cite{Defboot}:
\begin{equation}
T_{-}(\theta+\frac{i\pi}{3})T_{-}(\theta-\frac{i\pi}{3})=T_{-}(\theta)\label{eq:LYTboot}\end{equation}
Using this relation together with the defect unitarity and defect
crossing symmetry (\ref{Defprop}) we can fix the transmission factor
as \begin{equation}
T_{-}(\theta)=[b+1][b-1]\quad;\qquad[x]=i\frac{\sinh(\frac{\theta}{2}+i\frac{\pi x}{12})}{\sinh(\frac{\theta}{2}+i\frac{\pi x}{12}-i\frac{\pi}{2})}\label{eq:YLdefsol}\end{equation}
(Actually the inverse of the solution is also a solution but the two
are related by the $b\to6+b$ transformation). The defect energy,
as in the sinh-Gordon case, can be obtained as \[
\epsilon_{\mathrm{def}}=\epsilon_{\mathrm{bdry}}-\epsilon_{\mathrm{bdry}}^{\mathbb{I}}=m\sin\frac{b\pi}{6}\]
We are going to recover this expression from the UV analysis of defect
TBA in section 5. We also note that the defect with parameter $b=3$
behaves as a standing particle both from the energy and from the scattering
point of view.

\section{Peturbative calculations}

In this section we check the exact solution of the ShG defect system
in the free/classical ($b\to0$) limit, and develop a systematic perturbative
expansion.\emph{ }The parameter $b^{2}$ plays the same role as $\hbar$
which can be seen by scaling it out from the Lagrangian via $\Phi\to b\Phi$.
Since the classical groundstate $\Phi=0$ is invariant under this
scaling, the $b^{2}\to0$ limit corresponds both to the free and also
to the classical limit. Moreover, the perturbative expansion in $b^{2}$
is equivalent both to the loop expansion and to the semi-classical
approximation.

\subsection{Classical/free limit}

As a first step we identify the $b\to0$ limit of the defect Lagrangian
(\ref{eq:ShGdefL}) as: \begin{eqnarray*}
\mathcal{L} & = & \Theta(-x)\left[\frac{1}{2}(\partial_{\mu}\Phi_{-})^{2}-\frac{m_{\mathrm{cl}}^{2}}{2}\Phi_{-}^{2}\right]+\Theta(x)\left[\frac{1}{2}(\partial_{\mu}\Phi_{+})^{2}-\frac{m_{\mathrm{cl}}^{2}}{2}\Phi_{+}^{2}\right]\\
 &  & -\frac{\delta(x)}{2}\left(\Phi_{+}\dot{\Phi}_{-}-\Phi_{-}\dot{\Phi}_{+}+m_{\mathrm{cl}}\left[\cosh\mu\left(\Phi_{+}^{2}+\Phi_{-}^{2}\right)+2\sinh\mu\,\Phi_{+}\Phi_{-}\right]\right)\end{eqnarray*}
Then we expand the fields on the two sides of the defect in terms
of creation/annihilation operators \[
\Phi_{\pm}(x,t)=\int_{-\infty}^{\infty}\frac{dk}{2\pi}\frac{1}{2\omega(k)}\left(a_{\pm}(k)e^{ikx-i\omega(k)t}+a_{\pm}^{+}(k)e^{-ikx+i\omega(k)t}\right)\quad;\quad\omega(k)=\sqrt{k^{2}+m_{\mathrm{cl}}^{2}}\]
 where the $a,a^{+}$ operators are adjoint of each other with commutators:\[
[a_{\pm}(k),a_{\pm}^{+}(k^{'})]=2\pi2\omega(k)\delta(k-k^{'})\]
Imposing the defect condition (obtained by varying the action) at
the origin \begin{eqnarray*}
\pm\partial_{t}\Phi_{\pm}\mp\partial_{x}\Phi_{\mp} & = & m_{\mathrm{cl}}(\sinh\mu\,\Phi_{\pm}+\cosh\mu\,\Phi_{\mp})\end{eqnarray*}
we can connect the creation/annihilation operators as \[
a_{\pm}(\pm k)=T_{\mp}(k)a_{\mp}(\pm k)\quad;\qquad T_{\mp}(k)=-\frac{m_{\mathrm{cl}}\sinh\mu\mp i\omega(k)}{m_{\mathrm{cl}}\cosh\mu-ik}\quad;\quad k>0\]
This shows that the defect is purely transmitting, that is we do not
have any reflected wave. The transmission factor in the rapidity parametrization
($k=m_{\mathrm{cl}}\sinh\theta)$ can be written also in the following
form: \[
T_{-}(\theta)=-i\frac{\sinh(\frac{\theta}{2}-\frac{i\pi}{4}+\frac{\mu}{2})}{\sinh(\frac{\theta}{2}+\frac{i\pi}{4}+\frac{\mu}{2})}\]
Clearly it has exactly the same form as the exact quantum one (\ref{Defsol})
except the $B\vartheta\leftrightarrow\mu$ replacement. Having observed
this coincidence the authors in \cite{sgdef} suggested that they
might be the same $B\vartheta=\mu$. Using our defect UV-IR relation
(\ref{eq:UVIRdef}) we can perform the expansion: \begin{equation}
B\vartheta=\mu(1-\frac{b^{2}}{8\pi}+\dots)\label{eq:murenb2}\end{equation}
The term of first order shows that our exact solution is correct in
the classical limit, i.e. for $b\to0$. The term of second order shows
that the $B\vartheta=\mu$ relation suggested in \cite{sgdef} is
not valid: $B\vartheta$ acquires nontrivial quantum correction. Since
the renormalization of the parameter $\mu$ is crucial to decide about
the two proposals we perform a perturbative check at order $b^{2}$.

\subsection{Perturbation theory}

As a first step we collect the free propagators. If the fields are
on the same side of the defect we have \vspace{1cm}

\begin{tabular}{cc}
\multicolumn{1}{l}{\noindent $_{0}\langle0\vert T\left(\Phi_{\pm}(x,t)\Phi_{\pm}(x^{'},t^{'})\right)\vert0\rangle_{0}=\int\frac{d^{2}q}{(2\pi)^{2}}\frac{i}{q^{2}-m_{\mathrm{cl}}^{2}+i\epsilon}e^{iq(y-y^{'})}=G_{\pm}^{\pm}(y,y')$ } & \noindent \vspace{-1.5cm}\tabularnewline
 & \noindent \includegraphics[height=2cm]{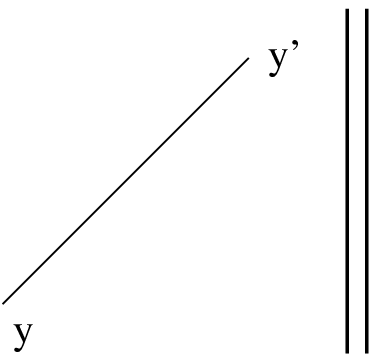}\tabularnewline
\end{tabular}

\noindent where $q=(k,\omega)$ and $y=(x,t)$. The absence of an
$e^{ik(x+x^{'})}$ term shows the absence of reflection. The other
two point functions are \vspace{1cm}

\begin{tabular}{cc}
\multicolumn{1}{l}{\noindent $_{0}\langle0\vert T\left(\Phi_{\mp}(x,t)\Phi_{\pm}(x^{'},t^{'})\right)\vert0\rangle_{0}=\int\frac{d^{2}q}{(2\pi)^{2}}\frac{i}{q^{2}-m_{\mathrm{cl}}^{2}+i\epsilon}T_{\pm}(\omega,k)e^{iq(y-y^{'})}=G_{\mp}^{\pm}(y,y')$ } & \noindent \vspace{-1.5cm}\tabularnewline
 & \noindent \includegraphics[height=2cm]{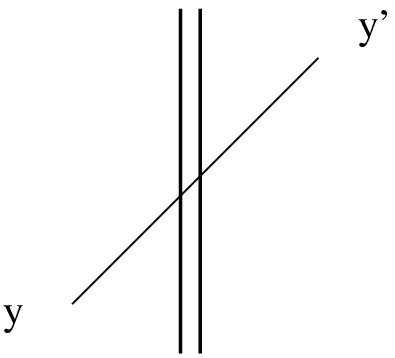}\tabularnewline
\end{tabular}

\noindent where \[
T_{\pm}(\omega,k)=-\frac{m_{\mathrm{cl}}\sinh\mu\pm i\omega}{m_{\mathrm{cl}}\cosh\mu-ik}\]
In the final equations we used the fact that the $\omega$ contour
can be closed on the upper/lower half plane. As it was shown in \cite{BRF}
the reflection factor can be read off from the $\langle T(\Phi_{\pm}\Phi_{\pm})\rangle$propagator
of the fields. The defect/boundary equivalence \cite{DefBound} then
implies that the transmission factor can be read off from the $\langle T(\Phi_{\mp}\Phi_{\pm})\rangle$
propagator. 

The perturbation at order $b^{2}$ follows from (\ref{ShGbulkL}):
\begin{eqnarray*}
\delta\mathcal{L} & = & -\Theta(-\zeta)\left[\frac{m_{\mathrm{cl}}^{2}b^{2}}{4!}\Phi_{-}^{4}\right]-\Theta(\zeta)\left[\frac{m_{\mathrm{cl}}^{2}b^{2}}{4!}\Phi_{+}^{4}\right]\\
 &  & -\delta(\zeta)\frac{m_{\mathrm{cl}}b^{2}}{4\cdot4!}\left[\cosh\mu\left(\Phi_{+}^{4}+6\Phi_{+}^{2}\Phi_{-}^{2}+\Phi_{-}^{4}\right)+4\sinh\mu\,\Phi_{+}\Phi_{-}\left(\Phi_{+}^{2}+\Phi_{-}^{2}\right)\right]\end{eqnarray*}
We calculate the propagators upto first order in $b^{2}$: \[
\langle0\vert T\left(\Phi_{\mp}(x,t)\Phi_{\pm}(x^{'},t^{'})\right)\vert0\rangle=\,\,_{0}\langle0\vert T\left(\Phi_{\mp}(x,t)\Phi_{\pm}(x^{'},t^{'})(1-i\int d\zeta\int d\tau\,\delta\mathcal{L}+\dots)\right)\vert0\rangle_{0}\]
Using Wick's theorem we obtain the contribution of the following diagrams: 

We have two bulk diagrams presented on Figure 1: 

\begin{center}
\begin{figure}[!h]
\begin{centering}
\includegraphics[height=2.5cm]{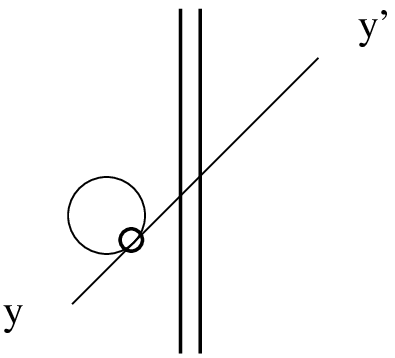}\hspace{2cm}\includegraphics[height=2.5cm]{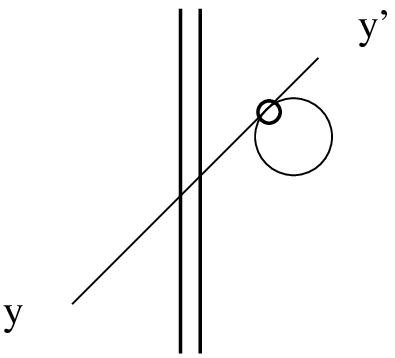}
\par\end{centering}

\caption{The two bulk diagrams with prefactor $\frac{m_{\mathrm{cl}}^{2}b^{2}}{2}$}

\end{figure}

\par\end{center}

\noindent where the bulk interaction point, denoted by an empty circle,
represents $z=(\zeta,\tau)$ and we have to integrate over the whole
left/right space-time. Thus the contribution of the first diagram
is \[
\frac{m_{\mathrm{cl}}^{2}b^{2}}{2}\int_{-\infty}^{\infty}d\tau\int_{-\infty}^{0}dz\, G_{-}^{-}(y,z)G_{-}^{-}(z,z)G_{-}^{+}(z,y')\]
Clearly $G_{-}^{-}(z,z)$ is divergent and we have to regularize it
by introducing a cutoff $\Lambda$: \[
G_{-}^{-}(z,z)=\int\frac{d^{2}q}{(2\pi)^{2}}\frac{i}{q^{2}-m_{\mathrm{cl}}^{2}+i\epsilon}=\int_{0}^{\Lambda}\frac{1}{\sqrt{k^{2}+m_{\mathrm{cl}}^{2}}}\frac{dk}{2\pi}=\Delta(m_{\mathrm{cl}})\]
Its contribution can be absorbed into the renormalization of the mass
parameter $m_{\mathrm{cl}}^{2}\to m_{\mathrm{cl}}^{2}-m_{\mathrm{cl}}^{2}\frac{b^{2}}{2}\Delta(m_{\mathrm{cl}})$
which results in extra counter term diagrams presented on Figure 2:

\noindent \begin{center}
\begin{figure}[!h]
\noindent \begin{centering}
\includegraphics[height=2.5cm]{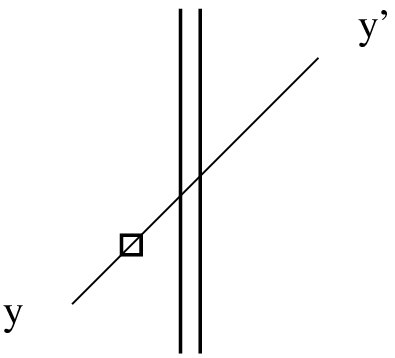}\hspace{2cm}\includegraphics[height=2.5cm]{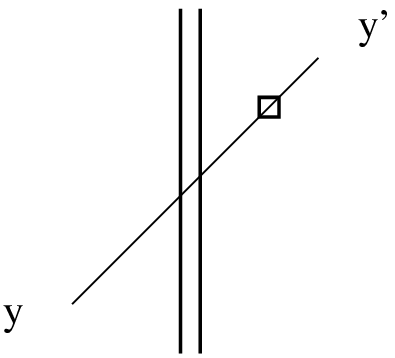}
\par\end{centering}

\caption{Bulk counter-term diagrams with prefactor $-m_{\mathrm{cl}}^{2}\frac{b^{2}}{2}\Delta(m_{\mathrm{cl}})$}

\end{figure}
\vspace{-1cm}
\par\end{center}

\noindent The contributions from the defect terms can be grouped in
two sets of diagrams. The first contains the same divergent loop integral
and consists of those on Figure 3: 

\begin{center}
\begin{figure}[H]
\begin{centering}
\includegraphics[height=2.5cm]{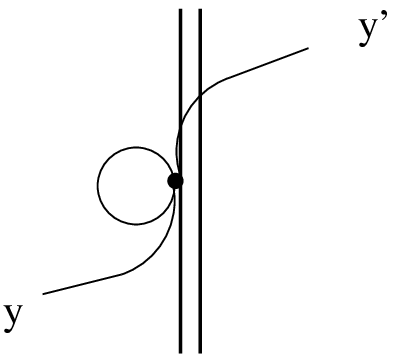}\includegraphics[height=2.5cm]{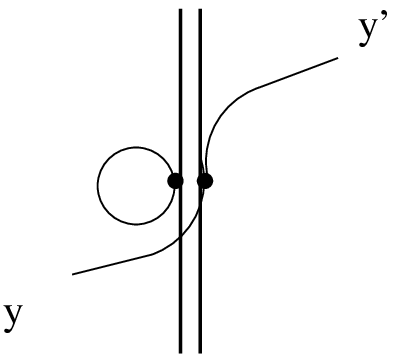}\includegraphics[height=2.5cm]{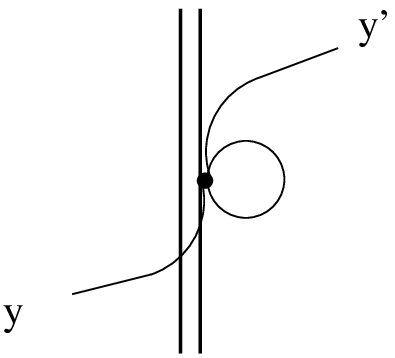}\includegraphics[height=2.5cm]{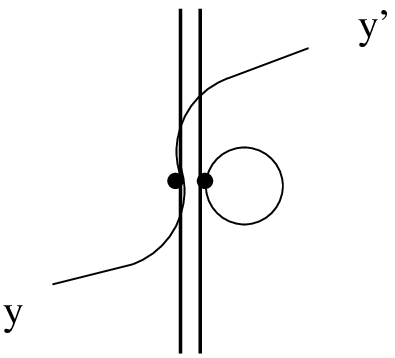}
\par\end{centering}

\caption{Divergent even defect loop diagrams with prefactor $\frac{m_{\mathrm{cl}}b^{2}}{8}\cosh\mu$ }

\end{figure}
\vspace{-1cm}
\par\end{center}

\noindent where the interaction vertex is even together with the odd
diagrams presented on Figure 4:

\begin{center}
\begin{figure}[!h]
\begin{centering}
\includegraphics[height=2.5cm]{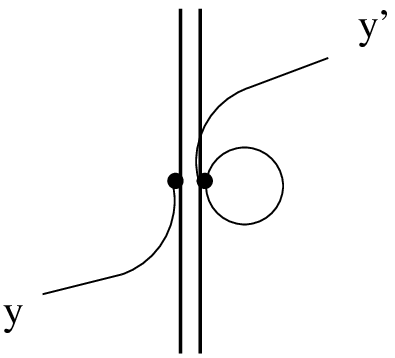}\includegraphics[height=2.5cm]{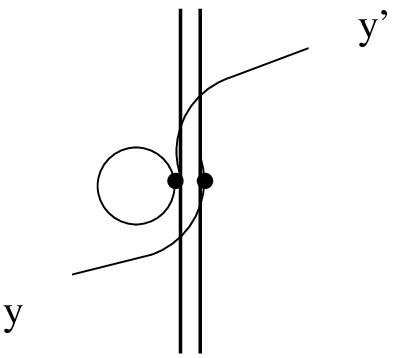}\includegraphics[height=2.5cm]{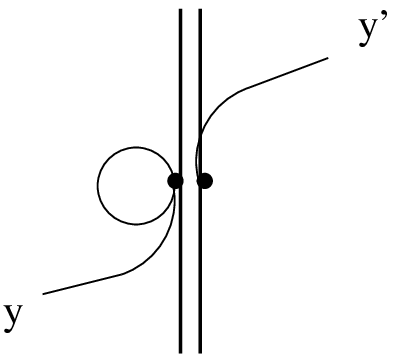}\includegraphics[height=2.5cm]{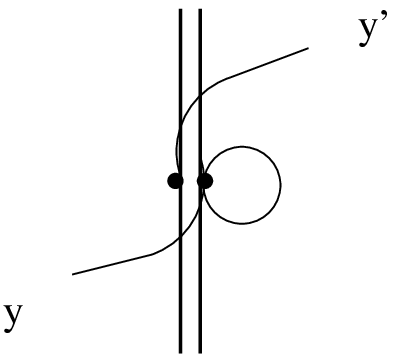}
\par\end{centering}

\caption{Divergent odd defect loop diagrams with prefactor $\frac{m_{\mathrm{cl}}b^{2}}{8}\sinh\mu$}

\end{figure}
\vspace{-1cm}
\par\end{center}

\noindent We have to integrate in time $\tau$ over the real axis
and the left/right part of the defect represents the contraction with
the operators $\Phi_{-}$ and $\Phi_{+}$, respectively. They all
contain the divergent and regularized $\Delta(m_{\mathrm{cl}})$ loop
integral which can be absorbed into the renormalization of the defect
parameter $m_{\mathrm{cl}}\to m_{\mathrm{cl}}-m_{\mathrm{cl}}\frac{b^{2}}{4}\Delta(m_{\mathrm{cl}})$,
which is consistent with the bulk renormalization. The resulting counter-terms
produce the diagrams on Figure 5:

\begin{figure}[H]
\begin{centering}
\includegraphics[height=2.5cm]{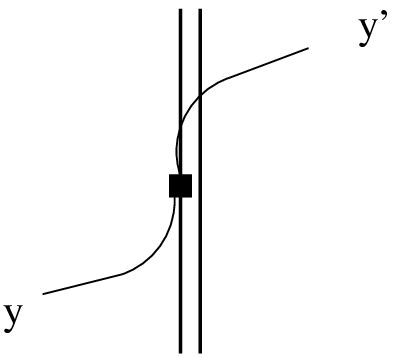}\includegraphics[height=2.5cm]{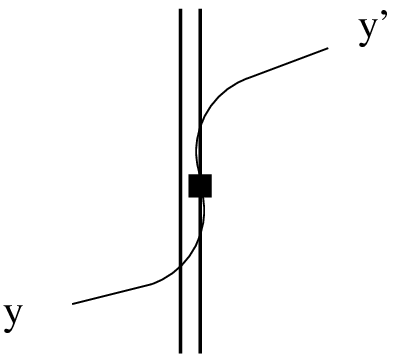}\hspace{3cm}\includegraphics[height=2.5cm]{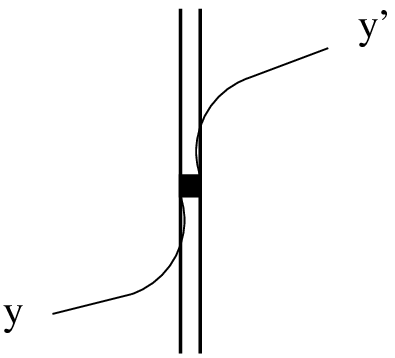}\includegraphics[height=2.5cm]{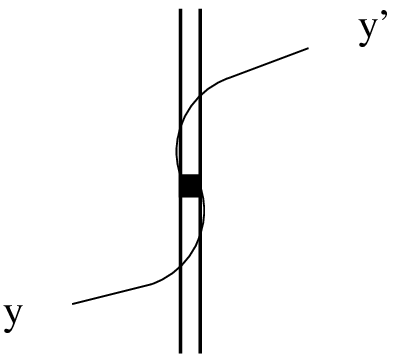} 
\par\end{centering}

\caption{Defect counter terms with prefactors $-m_{\mathrm{cl}}\frac{b^{2}}{4}\Delta(m_{\mathrm{cl}})\cosh\mu$
and $-m_{\mathrm{cl}}\frac{b^{2}}{4}\Delta(m_{\mathrm{cl}})\sinh\mu$,
respectively }

\end{figure}

\noindent The fact that all these singularities can be absorbed into
the renormalization of $m_{\mathrm{cl}}$ is a nontrivial statement,
since we have eight divergent diagrams having different propagators
on the outer legs those we canceled just by renormalizing one single
parameter in the original Lagrangian. The form of the renormalized
Lagrangian is the same as the original one thus the integrable/topological
nature of the defect is not spoiled by quantum effects, there is no
anomaly. Observe also that the bulk mass term $m_{\mathrm{cl}}^{2}$
and the boundary term $m_{\mathrm{cl}}$ renormalizes the same way
so the bulk is the square of the other. 

The last group of the diagrams is the one which really contributes
to the transmission factor. They are presented on Figure 6: 

\begin{figure}[H]
\begin{centering}
\includegraphics[height=2.5cm]{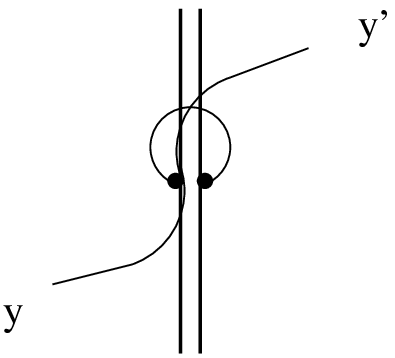}\includegraphics[height=2.5cm]{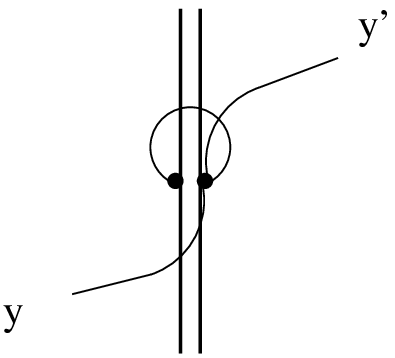}\hspace{3cm}\includegraphics[height=2.5cm]{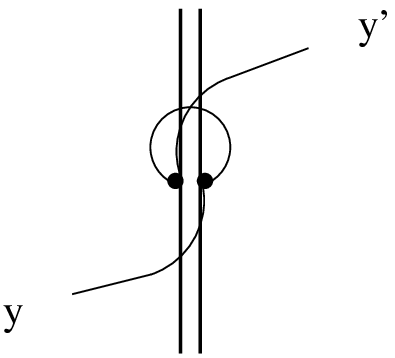}\includegraphics[height=2.5cm]{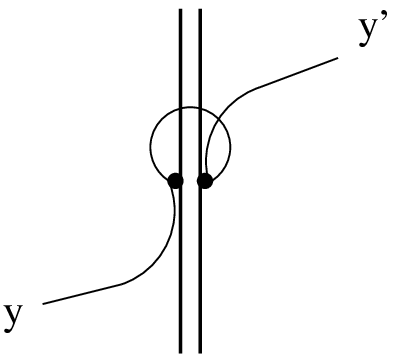}
\par\end{centering}

\caption{Defect diagrams contributing to the transmission factor. They have
prefactors $\frac{m_{\mathrm{cl}}b^{2}}{4}\sinh\mu$ and $\frac{m_{\mathrm{cl}}b^{2}}{4}\cosh\mu$,
respectively}

\end{figure}

\noindent The contribution of the first diagram is \[
\frac{m_{\mathrm{cl}}b^{2}}{4}\sinh\mu\int_{-\infty}^{\infty}d\tau\, G_{-}^{-}(y,z)G_{-}^{+}(z,z)G_{-}^{+}(z,y')\]
Each of the terms on Figure 6 contains the finite contribution of
the propagator \[
G_{\pm}^{\mp}(z,z)=\int\frac{d^{2}q}{(2\pi)^{2}}\frac{i}{q^{2}-m_{\mathrm{cl}}^{2}+i\epsilon}T_{\pm}(\omega,k)=-\frac{\mu}{2\pi}\]
This term together with the prefactor can be interpreted as the finite
renormalization of the parameter $\mu\to\mu+\delta\mu$, where $\delta\mu=-\frac{\mu b^{2}}{8\pi}$.
Summing up all the contributions and taking into account the different
transmission factor dependent contributions on the outer legs we obtain
the correction of order $b^{2}$ to the transmission factor as \[
T_{-}(\theta,B\vartheta(\mu))=T_{-}(\theta,\mu)+\frac{\mu b^{2}}{8\pi}\frac{1}{1-i\sinh(\theta+\mu)}+\mathcal{O}(b^{4})\]
which is in complete agreement with (\ref{eq:murenb2}). Thus we confirmed
the renormalization of the parameter $\mu$, but we have not checked
the renormalization of the parameter $M$ which has not shown up at
this order. We suspect, however, that in this perturbative scheme
the boundary parameter $m_{\mathrm{cl}}$ renormalizes as the square-root
of the $m_{\mathrm{cl}}^{2}$ term and only $\mu$ renormalizes as
$B\vartheta$. It would be interesting to perform a two-loop perturbative
calculation to decide about the renormalization of $M_{\mathrm{cl}}$
in the perturbative scheme. 

We note that we have also performed a perturbative calculation of
order $b^{2}$ of the reflection factor, which can be extracted from
$G_{+}^{+}(y,y')$, and confirmed the absence of reflection at this
order. In \cite{Patrick} the form of the sine-Gordon Lagrangian was
fixed at order $b^{6}$ by demanding the absence of particle creation.
Following a similar line it would be tempting to see how the absence
of reflection restricts the form of the defect potential in perturbation
theory.

\section{Defect thermodynamic Bethe ansatz}

In this section we would like to check the consistency between the
transmission factors and defect energies. In doing so we derive a
DTBA to describe the ground state energy of a purely transmitting
diagonal integrable defect on the circle of perimeter $L$. In boundary
and defect systems there are two inequivalent ways to derive TBA equations.
We can either focus on the groundstate energy or on the so-called
$g$-factors which is related to the finite volume normalization of
defect/boundary states. Both BTBA equations have been analyzed in
\cite{LMSS} although the $g$-function type required further refinement
\cite{gBTBA}. Some details of the $g$-function type DTBA can be
found in \cite{gDTBA} and references therein. Here we consider the
groundstate DTBA in the periodic setting as opposed to the strip geometry
analyzed in \cite{DefBound}.

\subsection{Derivation of DTBA}

In order to derive the DTBA equation for the ground state energy we
compactify the time-like direction with period $R$ and calculate
the partition function in two inequivalent ways by changing the role
of the space and time coordinates.

In the original description the defect is located in space as drawn
in Figure 7:

\begin{center}
\begin{figure}[!h]
\begin{centering}
\includegraphics[height=5cm]{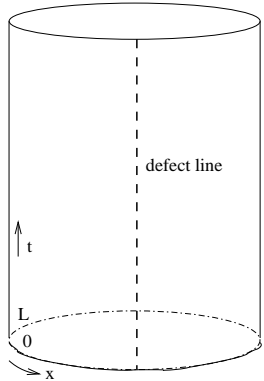}
\par\end{centering}

\caption{Defect is located in space}

\end{figure}

\par\end{center}

\noindent By taking the $R\to\infty$ limit the groundstate energy
can be extracted from the partition function as \[
\lim_{R\to\infty}Z(L,R)=\lim_{R\to\infty}Tr\left(e^{-H(L)R}\right)=e^{-E_{0}(L)R}+\dots\]
In the alternative description when the role of time and space is
exchanged as shown on Figure 8:

\begin{center}
\begin{figure}[!h]
\begin{centering}
\includegraphics[width=5cm]{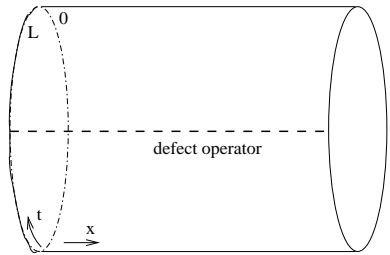}
\par\end{centering}

\caption{Defect is located in time acting as a defect operator}

\end{figure}

\par\end{center}

\noindent the defect becomes an operator of the form \cite{DefBound}
\[
D=\exp\left\{ \int_{-\infty}^{\infty}\frac{d\theta}{2\pi}T_{+}(\frac{i\pi}{2}-\theta)a^{+}(\theta)a(\theta)\right\} \]
 which acts on the Hilbert space of the periodic model, $\mathcal{H}$.
The partition function can be calculated as \begin{equation}
Z(L,R)=Tr\left(e^{-H(R)L}D\right)=\sum_{\vert n\rangle\in\mathcal{H}}\frac{\langle n\vert D\vert n\rangle e^{-E_{n}(R)L}}{\langle n\vert n\rangle}\label{eq:ZRL}\end{equation}
The Hilbert space consists of multi-particle states \[
\vert\theta_{1},\theta_{2},\dots,\theta_{n}\rangle=a^{+}(\theta_{1})a^{+}(\theta_{2})\dots a^{+}(\theta_{n})\vert0\rangle\quad;\qquad\theta_{1}>\theta_{2}>\dots>\theta_{n}\]
on which the defect operator collects nontrivial diagonal matrix elements
from \[
D\vert\theta_{1},\theta_{2},\dots,\theta_{n}\rangle=T_{+}(\frac{i\pi}{2}-\theta_{1})T_{+}(\frac{i\pi}{2}-\theta_{2})\dots T_{+}(\frac{i\pi}{2}-\theta_{n})\vert\theta_{1},\theta_{2},\dots,\theta_{n}\rangle+\dots\]
while the energy operator acts as\[
H\vert\theta_{1},\theta_{2},\dots,\theta_{n}\rangle=\left(m\cosh(\theta_{1})+m\cosh(\theta_{2})+\dots+m\cosh(\theta_{n})\right)\vert\theta_{1},\theta_{2},\dots,\theta_{n}\rangle\]
We can introduce another energy operator via $\hat{H}=H-\frac{1}{L}\log D$
such that the partition function can be written as \[
Z(L,R)=Tr\left(e^{-H(R)L}D\right)=Tr\left(e^{-\hat{H}(R)L}\right)=\sum_{\vert n\rangle\in\mathcal{H}}e^{-\hat{E}_{n}(R)L}\]
This partition function can be calculated in the $R\to\infty$ limit
by standard saddle point approximation taking into account the scattering
of the particles. The calculation follows the usual route  of TBA
calculations, but now the kinetic term is shifted $m\cosh\theta\to m\cosh\theta-\frac{1}{L}\log T_{+}(\frac{i\pi}{2}-\theta)$.
As a consequence we obtain the following DTBA equations for the pseudo
energy \begin{equation}
\tilde{\epsilon}(\theta)=mL\cosh\theta-\log T_{+}(\frac{i\pi}{2}-\theta)-\int_{-\infty}^{\infty}\frac{d\theta^{'}}{2\pi}\phi(\theta-\theta^{'})\log(1+e^{-\tilde{\epsilon}(\theta^{'})})\label{eq:DTBA1}\end{equation}
where $\phi(\theta)=-i\frac{d}{d\theta}\log S(\theta)$. Once $\tilde{\epsilon}$
is known the ground state energy can be expressed as \[
E_{0}(L)=-m\int_{-\infty}^{\infty}\frac{d\theta}{2\pi}\,\cosh\theta\,\log(1+e^{-\tilde{\epsilon}(\theta)})\]
 Here we do not have to shift the $\cosh\theta$ term since it comes
from the derivative of the momentum ($\sinh\theta$), which appears
in the quantization condition. For the result in this generality see
e.g. \cite{O(1)}. 

Alternatively, we can redefine the pseudo energy as $\tilde{\epsilon}(\theta)=\epsilon(\theta)-\log T_{+}(\frac{i\pi}{2}-\theta)$
to obtain \[
\epsilon(\theta)=mL\cosh\theta-\int_{-\infty}^{\infty}\frac{d\theta^{'}}{2\pi}\phi(\theta-\theta^{'})\log\left(1+T_{+}(\frac{i\pi}{2}-\theta^{'})e^{-\epsilon(\theta^{'})}\right)\]
from which the ground state energy turns out to be \[
E_{0}(L)=-m\int_{-\infty}^{\infty}\frac{d\theta}{2\pi}\,\cosh\theta\,\log\left(1+T_{+}(\frac{i\pi}{2}-\theta)e^{-\epsilon(\theta)}\right)\]
The ground state energy is real which can be easily seen form (\ref{Defprop})
since $T_{+}^{*}(\frac{i\pi}{2}-\theta)=T_{+}(\frac{i\pi}{2}+\theta)$.
As a simple consistency check we can see that the DTBA equation for
the trivial defect, $T_{+}=1$, reduces to the periodic TBA equation
\cite{TBA}. We analyze the large and small volume limits separately
in the next two subsections.

\subsection{L\"uscher type correction in defect systems}

If the volume $L$ is large then $\epsilon(\theta)\cong mL\cosh\theta$
is large and we can expand the logarithm to obtain\[
E_{0}(L)=-m\int_{-\infty}^{\infty}\frac{d\theta}{2\pi}\,\cosh\theta\,\, T_{+}(\frac{i\pi}{2}-\theta)e^{-mL\cosh\theta}+O(e^{-2mL})\]
This result can be calculated directly from (\ref{eq:ZRL}) by taking
the large $L$ limit there. In that case, however, only the one-particle
transmission term contributes which is universal for any quantum field
theory. The result obtained is the analogue of the boundary L\"uscher
type correction to the ground state energy \cite{BLusch} and is valid
in any theory even in non-integrable ones.

\subsection{Defect energy}

Here we analyze the small volume behavior of the ground state energy.
Its normalization depends on the scheme in which the quantum field
theory is defined. If we would like to compare the DTBA normalization
to that of a perturbed defect conformal field theory, in which the
perturbing operators have dimensions $h$ on the defect and $(h,h)$
in the bulk, then we have: \[
E_{0}(L)=-\epsilon_{\mathrm{def}}-\epsilon_{\mathrm{bulk}}L+\frac{2\pi}{L}\sum_{n=0}^{\infty}c_{n}\,{l}^{n(1-h)}\quad;\qquad l=mL\]
Only the perturbative terms, $c_{n}$, are present in a perturbed
rational defect CFT. (In non-rational CFT-s, like the UV limit of
the boundary sinh-Gordon theory, we expect terms with logarithmic
behaviour, see \cite{BRZ} for the details). By calculating the small
volume limit of $E_{0}(L)$ from DTBA  $\epsilon_{\mathrm{def}}$
and $\epsilon_{\mathrm{bulk}}$ can be extracted exactly. The computation
is analogous to the boundary one \cite{BUVIR,BYLTBA}, so we sketch
only here. In the $L\to0$ limit the solution for $\tilde{\epsilon}$
in (\ref{eq:DTBA1}) develops two kink regions around $\theta=\pm\log\frac{2}{l}$
and a breather region around the origin. The behaviour of the solutions
are determined by the $\theta\to\pm\infty$ asymptotics of the integral
kernel and defect source term: \[
\phi(\theta)=Ce^{-\vert\theta\vert}+O(e^{-2\vert\theta\vert})\quad;\quad\log(T_{+}(\frac{i\pi}{2}-\theta))=A_{\pm}e^{\mp\theta}+O(e^{\mp2\theta})\qquad\mathrm{as}\quad\theta\to\pm\infty\]
The two kink functions are responsible for the terms giving the central
charge and the bulk energy constant, while the central/breather part
gives the defect energy in the following form: \begin{equation}
\epsilon_{\mathrm{bulk}}=\frac{m^{2}}{2C}\quad;\qquad\epsilon_{\mathrm{def}}=-\frac{m(A_{+}+A_{-})}{2C}\label{eq:becdefe}\end{equation}
We note that the kink type behaviour does not exists for the whole
parameter range of $C$ and $A_{\pm}$. The results is understood
that we analytically continued it from a range where the calculation
is reliable. 

Let us concretes the result for the two cases in question. In the
sinh-Gordon model \[
C=4\sinh\pi B\quad;\qquad A_{\pm}=-2e^{\mp B\vartheta}\]
so using (\ref{eq:becdefe}) we recover (\ref{eq:becshg}) and (\ref{eq:Edef}). 

In the Lee-Yang case we have \[
C=-2\sqrt{3}\quad;\qquad A_{\pm}=\mp2i(e^{\pm i\pi\frac{b+1}{6}}+e^{\pm i\pi\frac{b-1}{6}})\]
Plugging these expressions back to (\ref{eq:becdefe}) the results
confirms the bulk energy density and the defect energy.

We emphasize that the agreement obtained in the two cases confirm
the solutions on one side and the DTBA equation on the other. 

The other perturbative coefficients $c_{n}$ can be calculated from
the DTBA only numerically. In the Lee-Yang case, however, one can
gain further analytical information. One has to define $Y(\theta)=e^{-\epsilon(\theta)}$
and to show from (\ref{eq:LYTboot}) that it satisfies the Lee-Yang
$Y$-system relation\[
Y(\theta-\frac{i\pi}{3})Y(\theta+\frac{i\pi}{3})=1+Y(\theta)\]
from which the $Y(\theta)=Y(\theta+\frac{5i\pi}{3})$ periodicity
follows. Similarly to the boundary case this gives the exponent of
the perturbative expansion to be $\frac{6}{5}$ showing that the dimension
of the perturbing operator is $h=-\frac{1}{5}$.\emph{ }

\section{Defect bound-states and bootstrap closure}

In this section we analyze the analytic structure of the transmission
factors for the whole range of their parameters. Since the sinh-Gordon
transmission factors (\ref{Defsol}) never have poles in the physical
strip we focus on the Lee-Yang model only. Recall that in our convention
the physical strip of the transmission factors $T_{\mp}(\theta)$
are $\Im m\theta\in[0,\frac{\pi}{2}]$.

\subsection{Pole analysis on the ground-state defect}

We analyze the pole structure of both \[
T_{-}(\theta)=[b+1][b-1]\quad\mbox{and}\quad T_{+}(\theta)=[5-b][-5-b]\]
as the function of the parameter $b$, simultaneously. We note that
by folding the theory to a boundary one (with two particles) we could
analyze its bootstrap in the usual boundary formulation. Here, however,
we present the results in the defect language since the diagrams are
more clear-cut. (The whole procedure, going from the reflection factor
to the defect transmission factor, can be interpreted as taking a
sort of square root of the boundary theory. By closing the defect
bootstrap we would like to show that such a theory is indeed sensible.
Since on the $\mathbb{I}$ boundary there is no boundstate any pole
of the reflection factor appears either in $T_{-}$ or in $T_{+}$
so the bootstrap will be very similar to the boundary one \cite{BCT}). 

In determining the fundamental range of the parameter $b$ we can
see that $b\to b+12$ is a symmetry. Moreover $b\leftrightarrow6-b$
exchanges $T_{-}\leftrightarrow T_{+}$ so we can restrict ourselves
to the range $b\in[-3,3]$. We will see by analyzing the defect excited
states that the fundamental range is even smaller, only $b\in[-3,2]$,
as in the boundary case, since at $b=2$ the role of the ground-state
and the first excited state is exchanged. 

The poles and zeros of the transmission factor $T_{-}(\theta)$ are
\[
\textrm{poles of $T_{-}$ are at }\theta=-i\frac{\pi}{6}(b\pm5)\quad;\qquad\textrm{zeros of $T_{-}$ are at }\theta=-i\frac{\pi}{6}(b\pm1)\]
The analogous expressions for $T_{+}$ are \[
\textrm{poles of $T_{+}$ are at }\theta=i\frac{\pi}{6}(b\pm1)\quad;\qquad\textrm{zeros of $T_{+}$ are at }\theta=i\frac{\pi}{6}(b\pm5)\]
 They can be drawn as the function of the parameter $b$ as shown
on Figure 9:

\begin{center}
\begin{figure}[!h]
\begin{centering}
\includegraphics[width=6cm]{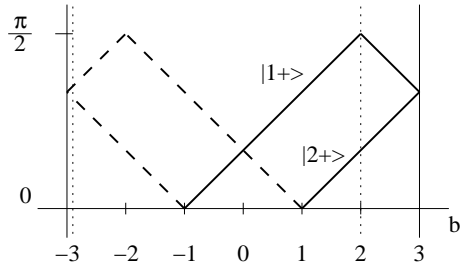}
\par\end{centering}

\caption{Poles and zeros of $T_{+}$ and $T_{-}$ as function of $b$ . Solid
lines correspond to poles, while dashed ones to zeroes. The dotted
lines show the fundamental range.}

\end{figure}

\par\end{center}

For $b\in[-1,2]$ there is a pole in the transmission factor $T_{+}$
at $\theta=iu=i\frac{\pi}{6}(b+1)$, for which we associate a defect
boundstate and denote it by $\vert1+\rangle$. Its energy is $m\cos\frac{\pi}{6}(b+1)$
and the corresponding excited transmission factor can be calculated
from the defect bootstrap equation shown on Figure 10:

\noindent \begin{center}
\begin{figure}[!h]
\noindent \begin{centering}
\includegraphics[height=4cm]{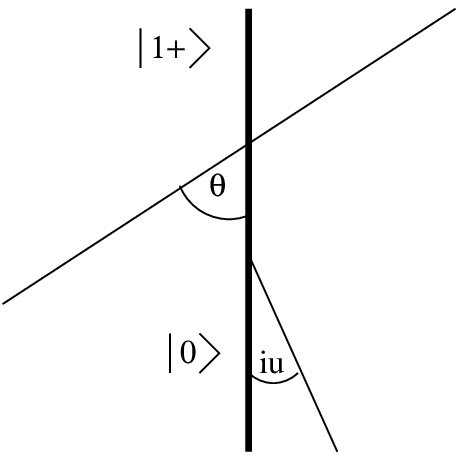}\hspace{2cm}\includegraphics[height=4cm]{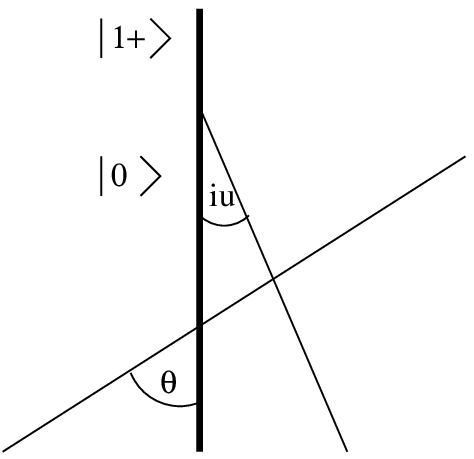}
\par\end{centering}

\caption{Defect bootstrap equations}

\end{figure}

\par\end{center}

\noindent \[
T_{-}^{\vert1+\rangle}(\theta)=T_{-}(\theta)S(\theta+iu)\]
From the defect crossing symmetry (\ref{Defprop}) we can calculate
$T_{+}^{\vert1+\rangle}(\theta)$ as \[
T_{+}^{\vert1+\rangle}(\theta)=T_{-}^{\vert1+\rangle}(i\pi-\theta)=T_{-}(i\pi-\theta)S(i\pi-\theta+iu)=T_{+}(\theta)S(\theta-iu)\]
which is consistent with the other bootstrap equation where the second
particle arrives from the right. The resulting transmission factors
are \[
T_{-}^{\vert1+\rangle}(\theta)=[b+1][b+3]\quad;\qquad T_{+}^{\vert1+\rangle}(\theta)=[5-b][3-b]\]
They are related to the groundstate ones as $T_{\pm}^{\vert1+\rangle}(b\to4-b,\theta)=T_{\mp}(b,\theta)$.
This symmetry together with the defect energies indicate that when
$b$ exceeds $2$ the role of the ground-state and the excited state
$\vert1+\rangle$ are exchanged. This confirms that the fundamental
range is indeed $b\in[-3,2]$. 

In the range $b\in[1,2]$ the transmission factor $T_{+}(\theta)$
has another pole at $\theta=i\frac{\pi}{6}(b-1)$ for which we associate
the defect boundstate $\vert2+\rangle$. It has energy $m\cos\frac{\pi}{6}(b-1)$
and transmission factor \[
T_{\pm}^{\vert2+\rangle}(\theta)=T_{\pm}(\theta)S(\theta\mp i\frac{\pi}{6}(b-1))\]
Now we turn to the pole analysis of excited defect states.

\subsection{Pole analysis on the exited defect state $\vert1+\rangle$}

The poles and zeros of the transmission factors on the state $\vert1+\rangle$
are indicated on Figure 11. 

\begin{center}
\begin{figure}[H]
\begin{centering}
\includegraphics[width=6cm]{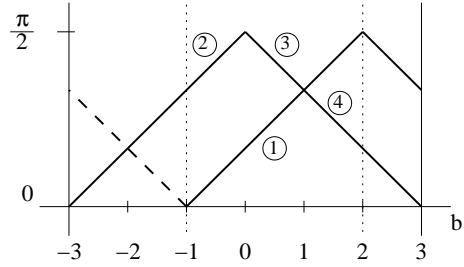}
\par\end{centering}

\caption{Poles and zeros on the $\vert1+\rangle$ defect excited state. Dotted
line shows the range where the excited state $\vert1+\rangle$ exists}

\end{figure}

\par\end{center}

\noindent The state exist in the $b\in[-1,2]$ domain so we have to
explain the poles in this range only. 

The pole of $T_{+}^{\vert1+\rangle}$ labeled by $1$ on Figure 11
is at the same location as the one which creates the excited state
itself, namely at $\theta=i\frac{\pi}{6}(b+1)$ in the full range
$b\in[-1,2]$. It can be explained by the first of the defect Coleman-Thun
diagrams on Figure 12 

\begin{center}
\begin{figure}[H]
\begin{centering}
\includegraphics[height=5cm]{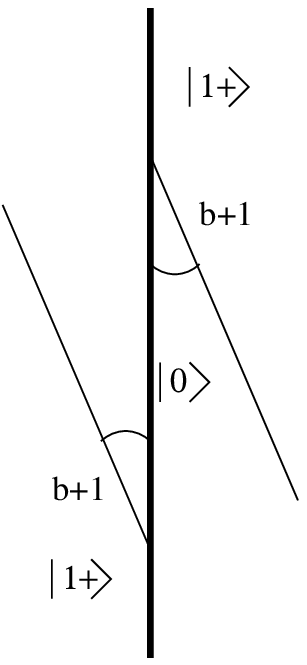}\hspace{1cm}\includegraphics[height=5cm]{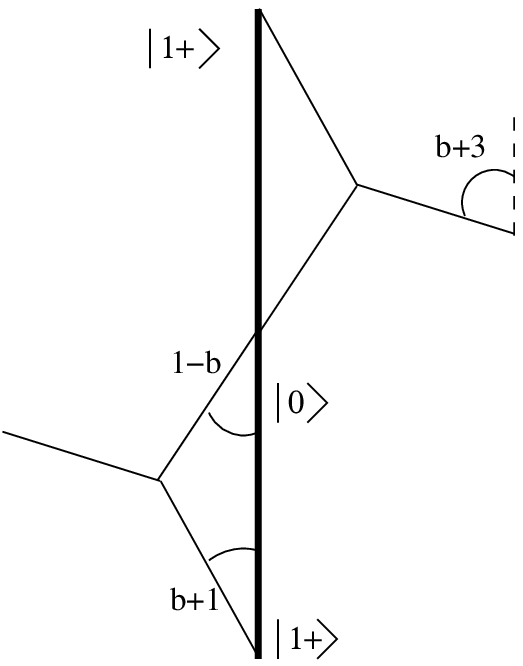}\hspace{1cm}\includegraphics[height=5cm]{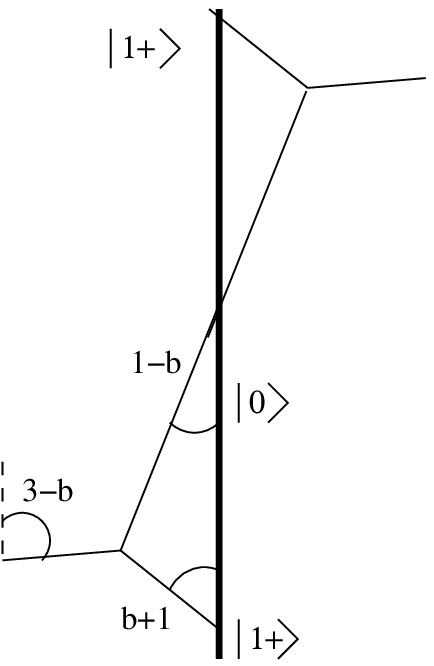}
\par\end{centering}

\caption{Defect Coleman-Thun diagrams for the $\vert1+\rangle$ state. The
angles are measured in units of $\frac{i\pi}{6}$ }

\end{figure}

\par\end{center}

The pole of $T_{+}^{\vert1+\rangle}$ labeled by $2$ on Figure 11
is at $\theta=i\frac{\pi}{6}(b+3)$ and can be explained by the second
diagram on Figure 12. Observe that by applying the Cutkosky rules
\cite{BBT} we would obtain a pole of second order but the transmission
factor $T_{-}$ has a first order zero at $\theta=i\frac{\pi}{6}(1-b)$
which, in this way, reduces the order of the pole to one.

The pole of $T_{-}^{\vert1+\rangle}$ labeled by $3$ on Figure 11
is at $\theta=i\frac{\pi}{6}(3-b)$. In the range $b\in[0,1]$ it
can be explained by the third diagram on Figure 12. Since the transmission
factor $T_{-}$ on the ground state has a zero at $\theta=i\frac{\pi}{6}(1-b)$
the order of the diagram is reduced to one again. In order for the
diagram to exist the particle has to travel towards the defect, that
is $1-b>0$. This explains the pole in the range $b\in[0,1]$. In
the range $b\in[1,2]$ the particle creates a defect boundstate which
is nothing but $\vert2+\rangle$. This can be seen both from the energy
of the excited state $m\cos\frac{\pi}{6}(b+1)+m\cos\frac{\pi}{6}(3-b)=m\cos\frac{\pi}{6}(b-1)$
and from the transmission factor. If the left particle creates a defect
boundstate at rapidity $\theta=iu$ then the excited states transmission
factors are $T_{\pm}^{ex}(\theta)=T_{\pm}(\theta)S(\theta\pm iu)$.
Now we can see from the bulk bootstrap equation that \[
T_{\pm}^{\vert1+\rangle}(\theta)S(\theta\pm i\frac{\pi}{6}(3-b))=T_{\pm}(\theta)S(\theta\mp i\frac{\pi}{6}(b+1))S(\theta\pm i\frac{\pi}{6}(3-b))=T_{\pm}(\theta)S(\theta\mp i\frac{\pi}{6}(b-1))=T_{\pm}^{\vert2+\rangle}(\theta)\]
that is the transmission factors also supports the identification.

\subsection{The pole analysis on the excited defect state $\vert2+\rangle$}

The defect boundstate labeled by $\vert2+\rangle$ has transmission
factor \[
T_{-}^{\vert2+\rangle}(\theta)=[b-1][b+1]^{2}[b+3]\quad;\qquad T_{+}^{\vert2+\rangle}(\theta)=[3-b][5-b]^{2}[7-b]\]
The singularity structure can be summarized as follows. 

\begin{center}
\begin{figure}[H]
\begin{centering}
\includegraphics[width=6cm]{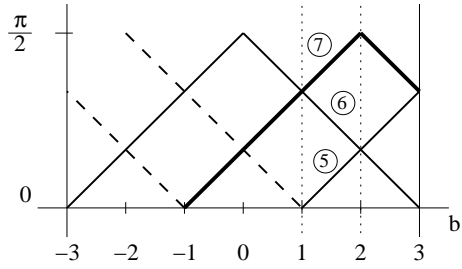}
\par\end{centering}

\caption{Singularity structure of the transmission factors on the $\vert2+\rangle$
state. Bold straight lines represent poles of second order. The relevant
interval where the boundstate $\vert2+\rangle$ exists is indicated
by dotted lines. }

\end{figure}

\par\end{center}

The pole labeled by $5$ on Figure 13 is in $T_{+}^{\vert2+\rangle}(\theta)$
at $\theta=i\frac{\pi}{6}(b-1)$ and can be explained in the full
range $b\in[1,2]$ by the first diagram on Figure 12, if we replace
$\vert1+\rangle$ by $\vert2+\rangle$. 

The pole labeled by $6$ on Figure 13 is in $T_{-}^{\vert2+\rangle}(\theta)$
at $\theta=i\frac{\pi}{6}(3-b)$ and can be explained by a diagram
similar to the third one of Figure 12 in which the $\vert1+\rangle$
state is replaced by $\vert2+\rangle$ and the vacuum $\vert0\rangle$
is replaced by $\vert1+\rangle$. 

\begin{center}
\begin{figure}[H]
\begin{centering}
\includegraphics[height=5cm]{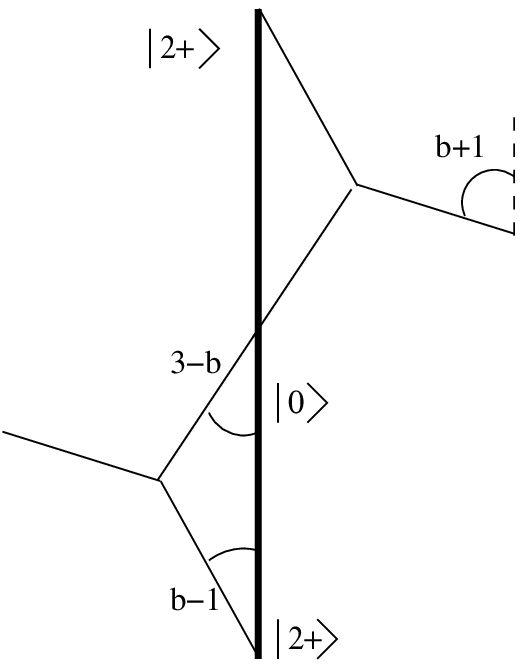}\hspace{2cm}\includegraphics[height=5cm]{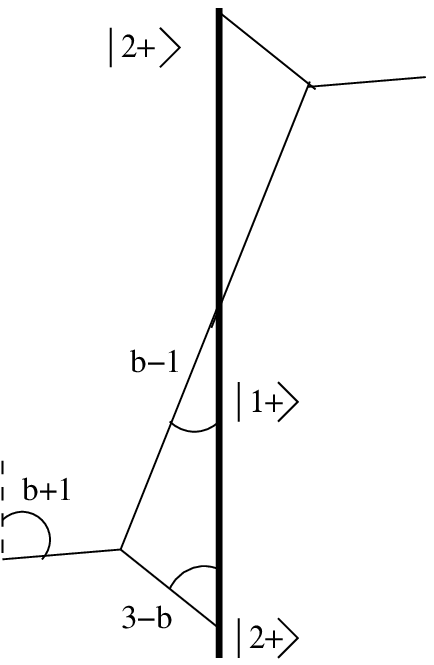}
\par\end{centering}

\caption{Defect Coleman-Thun diagrams for the excited state $\vert2+\rangle$}

\end{figure}

\par\end{center}

The pole labeled by $7$ on Figure 13 is a second order one in $T_{+}^{\vert2+\rangle}(\theta)$
at $\theta=i\frac{\pi}{6}(b+1)$ and can be explained by the two diagrams
on Figure 14. Clearly the transmission factor $T_{-}(\theta)$ does
not have zeros neither at $\theta=i\frac{\pi}{6}(3-b)$ nor at $\theta=i\frac{\pi}{6}(b-1)$
so the pole is of second order. 

By now we explained all the poles of all the transmission factors
of the ground and excited defect states. We used either the creation
of a new defect boundstate or presented the appropriate defect Coleman-Thun
diagram which was responsible for the singularity. By finishing this
procedure the spectrum become complete and we managed to define a
sensible defect theory. It would be nice to check these findings by
the defect truncated conformal space approach (TCSA).

\section{Conclusions}

We have demonstrated how the fusion idea can be used to solve topological
defects in the sinh-Gordon and Lee-Yang models. In the sinh-Gordon
case we determined the transmission factors and the defect energy
as a function of a bootstrap parameter whose relation to the Lagrangian
was also given. We checked these results in perturbation theory and
against the newly derived DTBA.

In the Lee-Yang case we determined the transmission factors together
with the defect energy and checked them in DTBA. For certain range
of the parameter the transmission factor admits poles in the physical
strip. We closed the defect bootstrap programme: we explained all
poles either by associating new defect boundstates or by giving the
appropriate defect Coleman-Thun mechanism both for the groundstate
and for excited defect states.

The relation obtained between the transmission parameter and that
of the Lagrangian in the sinh-Gordon theory can be analytically continued
to describe the analogues relation in the sine-Gordon theory. This
result also passes the test of first order perturbation theory and
together with the transmission factors obtained in \cite{KL,sgdef}
gives the complete solution of defect sine-Gordon model. We have checked
this solution by performing the fusing procedure on the solitonic
transmission factors. This is analogous to the dressing procedure
in the XXZ spin chain developed in \cite{XXZ}. 

The perturbation theory developed here can also be used in higher
rank affine Toda theories to connect the parameters of the transmission
factor of the bootstrap solution \cite{QAT} to the parameters of
their Lagrangians \cite{DAT}.

The derivation of the DTBA generalizes to any diagonal scattering
theory. A large and small volume analysis analogous to the one presented
in the paper will provide the leading finite size correction to the
groundstate energy and give the bulk/defect energies, respectively. 

In the present paper we were concerned with the bootstrap (IR) description
of our models. There is a need, however, to understand their UV behavior
which probably can be described by perturbed defect CFTs. To connect
these alternative descriptions we can use methods starting either
from the IR side, like DTBA, or starting from the UV side, like defect
TCSA. There are works in progress in both directions.

\subsection*{Acknowledgments}

ZB thanks Laszl\'o Palla, G\'abor Tak\'acs, Gerard Watts and Cristina
Zambon for the useful discussions. ZB was supported by a Bolyai Scholarship,
OTKA K60040 and the EC network {}``Superstring''.

\end{document}